\documentclass[12pt]{article}
\textheight
215 mm
\usepackage[dvips]{graphicx}
\usepackage{amsthm}
\usepackage{lipsum}
\usepackage{graphicx}
\usepackage[colorinlistoftodos]{todonotes}
\usepackage{amsthm}
\usepackage{latexsym}
\usepackage{enumitem}
\usepackage{cancel}
\usepackage{soul}
\usepackage{lineno}
\usepackage{amsmath}
\usepackage{amssymb}
\usepackage{hyperref}
\usepackage{cleveref}

\newtheorem{prop}{Claim}

\begin{document}
\sf
\begin{center}
   \vskip 2em
    {\LARGE \sf Stochastic curvature of enclosed semiflexible polymers}  
 \vskip 3em
 {\large \sf  Pavel Castro-Villarreal$^{a}$ and J. E. Ram\'irez$^{b,c}$ }\\
\em{ $^{a}$Facultad de Ciencias en F\'isica y Matem\'aticas,Universidad Aut\'onoma de Chiapas,\\ Carretera Emiliano Zapata, Km. 8, Rancho San Francisco, 29050\\ Tuxtla Guti\'errez, Chiapas, M\'exico
}\\
\em{$^{b}$Facultad de Ciencias F\'isico Matem\'aticas, Benem\'erita Universidad Aut\'onoma de Puebla,\\
Apartado Postal 165, 72000 Puebla, Pue., M\'exico}\\
\em{$^{c}$Departamento de F\'isica de Part\'iculas, Universidad de Santiago de Compostela, E-15782 Santiago de Compostela, Espa\~na}

\end{center}
 \vskip 1em



\begin{abstract}
The conformational states of a semiflexible polymer enclosed in a compact domain of typical size $a$ are studied as stochastic realizations of paths defined by the Frenet equations under the assumption that stochastic ``curvature'' satisfies a white noise fluctuation theorem. This approach allows us to derive the Hermans-Ullman equation, where we exploit a multipolar decomposition that allows us to show that the positional probability density function is well described by a Telegrapher's equation whenever $2a/\ell_{p}>1$, where $\ell_{p}$ is the persistence length. We also develop a Monte Carlo algorithm for use in computer simulations in order to study the conformational states in a compact domain. In addition,  the case of a semiflexible polymer enclosed in a square domain of side $a$ is presented as an explicit example of the formulated theory and algorithm. In this case, we show the existence of a polymer shape transition similar to the one found by Spakowitz and Wang [Phys. Rev. Lett. {\bf 91}, 2 (2003)] where in this case the critical persistence length is $\ell^{*}_{p}\simeq a/8$ such that the mean-square end-to-end distance exhibits an oscillating behavior for values $\ell_{p}>\ell^{*}_{p}$, whereas for $\ell_{p}<\ell^{*}_{p}$ it behaves monotonically increasing. 
\end{abstract}
\tableofcontents

\section{Introduction}\label{intro}

Semiflexible polymers is a term coined to understand a variety of physical systems that involve linear molecules. For instance, understanding the behaviors of such polymers serves as the basis to understand phenomena encountered in polymer industry, biotechnology, and mo\-le\-cu\-lar processes in living cells \cite{Fal-Sakaue2007}.  Indeed, biopolymers functionality are ruled by their conformation, which in turn is considerably modified in the geometrically confined or crowded environment inside the cell \cite{Fal-Koster2008, Fal-Reisner2005,Fal-Cifra2010, Fal-Benkova2017}.  Beyond the most prominent polymer example being DNA compaction in the nucleus or  viral DNA  packed in capsids \cite{Fal-Cifra2010, Fal-Locker2006}, there is also the important outstanding  example of DNA transcription and replication processes that are governed by the binding of specific proteins. These  mechanisms are strongly connected to polymer configuration \cite{Fal-Gowers2003,Fal-Broek2008,Fal-Ostermeir2010}. Furthermore, a wide range of biophysical processes is derived from  DNA constrained to a ring enclosure and more general topologies \cite{Fal-Witz2011}. 

On the one hand, motivated by the packaging and coiling problems mentioned above, Mondescu and Muthukumar (MM) studied in \cite{PDoi-Mondescu1998} the conformational states of an ideal Gaussian polymer \cite{PDoi-Doi1988book} wrapping different curved surfaces, where they presented theoretical predictions for the mean-square end-to-end distance.  Later on, Spakowitz and Wang (SW) in \cite{PSaito-Spakowitz2003} studied the conformational states of an ideal semiflexible polymer confined to a spherical surface based on the continuous Worm-Like Chain Model (WLC) \cite{PSaito-Saito1967}. Unlike the conformational states of the Gaussian polymer, SW found the existence of a shape transition from an ordered to a disordered phase, where polymer roughly looks like cooked spaghetti  and a random walk, respectively. Moreover, in the appropriate limit, the behavior of the semiflexible polymer reduces to the one of the Gaussian polymer in the spherical case. Subsequently, the MM and SW results were confirmed through computer simulations, where the validity regimes for each theory were established. Additionally,  as a consequence of the excluded volume effect a helical state was found in \cite{Fal-Cerda2005,Fal-Slosar2006} and a Tennis ball-like state in  \cite{Psim-Zhang2011}; these states are absent in both MM and SW theories. The problem becomes richer when short-range and long-range electrostatic interactions are considered since they can induce a Tennis ball-like state  which is predominant over the helical conformation in the case of long-range electrostatic interactions \cite{PSim-Angelescu2008}. A transition is also reported in a similar manner to that reported by SW  with slight corrections in the case of short-range interaction and more pronounced in the long-range ones. In the extreme limit of zero temperature, the conformational states are expected to be in the ordered phase. Indeed, by a variational principle consistent to the WLC model one can obtain an abundance of conformational states including those observed in the simulations mentioned above \cite{Psim-Guven2012,Psim-Lin2007} . On the other hand, the confinement can induce a transition from a circular polymer to a figure eight shape \cite{Fal-Ostermeir2010}; even when the conformation of the polymer is considered as a self-avoiding random walk, properties similar to that of a critical phenomenon are found \cite{Fal-Berger2006,Fal-Sakaue2006,Fal-Sakaue2007,Fal-Usatenko2017}. Confinement can also occur in the three-dimensional volume enclosed by rigid or soft surfaces \cite{Fal-Brochard2005, Fal-Cacciuto2006, Fal-Chen2007, Fal-Koster2008, PSaito-Morrison2009, Psim-Gao2014},  and in a crowded environment, for instance, modeled by a nanopost array \cite{Fal-Benkova2017}. 

Furthermore, two-dimensional confinement in closed flat spaces can result in a order-disorder shape transition \cite{Polcon-Liu2008} similar to that of SW. One advantage of confinement in the flat  two-dimensional case is that it can be compared with the experiment \cite{Fal-Moukhtar2007,Fal-Witz2011, Fal-Japaridze2017, Fal-Frielinghaus2013}. However, in the literature as far as we know, even in the semiflexible ideal chain  there is not a systematic study of the SW transition in a flat and bounded region. In this work, we present a theoretical and numerical analysis of the conformational states of an ideal semiflexible polymer in a compact two-dimensional flat space. First, we deduce the Hermans-Ullman (HU) equation \cite{PSaito-Hermans1952}  under the supposition that the conformational states correspond to stochastic realizations of paths defined by the Frenet equations and the assumption that stochastic ``curvature'' satisfies a fluctuation theorem given by a white noise distribution. This latter hypothesis is consistent with the continuous version of the WLC model as we will see below. Using the HU equation we shall perform a multipolar decomposition for the probability density function $P({\bf R}, \theta, {\bf R}^{\prime}, \theta^{\prime}, L)$  that gives the probability to find a polymer with length $L$ with endings ${\bf R}$  and  ${\bf R}^{\prime}$, and directions $\theta$ and $\theta^{\prime}$, respectively. This decomposition allows us to find a hierarchy of equations associated to the multipoles of $P({\bf R}, \theta, {\bf R}^{\prime}, \theta^{\prime}, L)$, namely, the positional  density distribution $\rho({\bf R}, {\bf R}^{\prime}, L)$, the dipolar distribution $\mathbb{P}^{i}\left({\bf R}, {\bf R}^{\prime}, L\right)$, the quadrupolar two-rank tensor distribution $\mathbb{Q}_{ij}({\bf R}, {\bf R}^{\prime}, L)$, and so on.  We shall show, for instance, that the positional density and the quadrupolar distributions are exactly related through the modified Telegrapher's Equation (MTE), 
\begin{eqnarray}
\frac{\partial^2\rho}{\partial s^2}+\frac{1}{2\ell_{p}}\frac{\partial\rho}{\partial s}=\frac{1}{2}\nabla^2\rho+\partial_{i}\partial_{j}\mathbb{Q}^{ij}.
\end{eqnarray}
In particular, using this equation and the traceless condition of  $\mathbb{Q}_{ij}$, we are going to verify the well known exact result of Kratky-Porod for a semiflexible polymer in a two-dimensional space \cite{PSaito-Kratky1949}.   Besides, we will show that as a consequence of the exponential decay of $\mathbb{Q}_{ij}$ we are going to define a regime where quadrupolar distribution can be neglected in the MTE. In addition, we shall explore the conformational states for a semiflexible chain enclosed by a bounded compact two-dimensional domain through the mean-square end-to-end distance. In particular, for a square domain we will show the existence of a shape transition order-disorder of the same nature as the one found by SW \cite{PSaito-Spakowitz2003}. Furthermore, we will develop a Monte Carlo algorithm for use in computer simulations in order to study the conformational states enclosed in a compact domain. Particularly, the algorithm shall be suited in the square domain which, additionally, will allow us to confirm the shape transition and validate the theoretical predictions.

This paper is organized as follows. In Sec. II, we present the stochastic version of the Frenet equations whose Fokker-Planck formalism  give us a derivation of  the Hermans-Ullman (HU) equation. In addition, we discuss a multipolar decomposition for the  HU equation.  In Sec. III, we provide an application of  the methods developed in Sec. II in order to study semiflexible polymer conformations enclosed in a compact domain. Particularly, we focus on  a square box domain. In Sec. IV, we present a  Monte Carlo algorithm to study the conformational states of a semiflexible polymer enclosed in a compact domain. In Sec. V,  we give the main results in  a square box domain and we provide a comparison with the theoretical predictions. In the final Sec. VI, we give our concluding remarks and perspectives on this work.

\section{Preliminary notation and semiflexible polymers  }\label{theory}

Let us consider a polymer on a two dimensional Euclidean space as a plane curve $\gamma$,  ${\bf R}: I\subset \mathbb{R}\to\mathbb{R}^2$, parametrized by an arc-length, $s$.  For each point $s\in I$, a Frenet dihedral can be defined whose vector basis corresponds to the set
 $\{{\bf T}(s), {\bf N}(s)\}$, consisting of the tangent vector ${\bf T}(s)={\bf R}^{\prime}(s)\equiv d{\bf R}/ds$  and the normal vector ${\bf N}(s)=\boldsymbol{\epsilon}{\bf T}(s)$, where $\boldsymbol{\epsilon}$ is a rotation by an angle of $\pi/2$. Note that the components of the rotation correspond to the Levi-Civita antisymmetric tensor in two dimensions.  Both are unit vectors ($\left|{\bf T}(s)\right|=\left|{\bf N}(s)\right|=1$), and by construction are orthogonal to each other. It is well known that along the points of the curve these vectors satisfy the Frenet equations, ${\bf T}^{\prime}(s)=\kappa(s){\bf N}(s)$ and ${\bf N}^{\prime}(s)=-\kappa(s){\bf T}(s)$,  where $\kappa(s)$ is the curvature of the curve \cite{Fal-Montiel2009}. 
 
In absence of thermal fluctuations, the conformations of the polymer are studied through different curve configurations determined by variational principles. For instance, one of the most successful models to des\-cribe  configurations of a semiflexible polymer is
\begin{eqnarray}
H[{\bf R}]=\frac{\alpha}{2}\int ds ~\kappa^2\left(s\right) , 
\label{funcional}
\end{eqnarray}
where $H[{\bf R}]$ is the bending energy, and $\alpha$ the bending rigidity modulus. This energy functional (\ref{funcional}) corresponds to the continuous form of the worm-like chain model (WLC) \cite{PSaito-Saito1967}. In a rather different context, a classical problem originally proposed by D. Bernoulli, later by L. Euler, between the XVIII and XIX centuries, consists of finding the family of curves $\{\gamma\}$ with a fixed length that minimizes the functional \eqref{funcional}. The solution to this problem is composed of those  curves whose curvature satisfies the differential equation $k^{\prime\prime}+\frac{1}{2}k^{3}-\frac{\lambda}{\alpha}\kappa=0$, where $\lambda$ is a Lagrange multiplier introduced to constrain the curve length \cite{Polcla-Miura2017}. This problem has been generalized to study elastic curves in manifolds \cite{Polcla-Singer2008, Psim-Guven2012, Polcla-Manning1987}, that are nowadays relevant to understand the problem of DNA packaging and the winding problem of DNA around histone octamers \cite{Fal-Hardin2012}. 


In what follows, we shall develop an unusual approach  that incorporates the thermal fluctuations in the study of semiflexible polymers described by the bending energy  \eqref{funcional}. 

\subsection{Stochastic Frenet Equations Approach}
In this section, we propose an approach to study conformational states of a semiflexible polymer immersed in a thermal reservoir and confined to a two-dimensional Euclidean space. We start by postulating that each conformational realization of any polymer on the plane is described by a stochastic path satisfying the stochastic Frenet equations defined by
\begin{subequations}
\label{ecsestom1}
\begin{align}
\label{ecsesto0}
\frac{d}{ds}{\bf R}\left(s\right)=&{\bf T}(s),\\
\frac{d}{ds}{\bf T}\left(s\right)=&{\kappa}(s){\boldsymbol{\epsilon}}{\bf T}(s),
\label{ecsesto}
\end{align}
\end{subequations}
where  ${\bf R}(s)$, ${\bf T}(s)$ and $\kappa(s)$ are now random variables. According to   this postulate, it can be show that $\left|{\bf T}(s)\right|$ is a constant that can be fixed to unit. 

In addition we postulate that $\kappa(s)$ for semiflexible polymers is a random variable, named here stochastic curvature, and is distributed according to the following probability density
\begin{eqnarray}
\mathcal{P}[\kappa]\mathcal{D}\kappa:=\frac{1}{Z}\exp\left[-\beta H\right]\mathcal{D}\kappa, 
\label{dft}
\end{eqnarray}
where $H$ is given by Eq.~\eqref{funcional},  $Z$ is an appropriate normalization constant, $\mathcal{D}\kappa$ is a functional measure, and $\beta=1/k_{B}T$   the inverse of the thermal energy $k_{B}T$, with $k_{B}$ and $T$ being the Boltzmann constant and the absolute temperature, respectively. It is also convenient to introduce the persistence length by $\ell_{p}=\beta\alpha$. Note that due to the Gaussian structure of the probability density \eqref{dft}, the stochastic curvature satisfies the following fluctuation theorem \cite{Fal-Zinn1996}
\begin{subequations}
\label{flucm1}
\begin{align}
\label{fluc0}
\left<\kappa(s)\kappa(s^{\prime})\right>=&\frac{1}{\ell_{p}}\delta(s-s^{\prime}),\\
\left<\kappa(s)\right>=&0. 
\label{fluc}
\end{align}
\end{subequations}
 
Since the polymer is confined to a plane and  ${\bf T}(s)$ is a unit vector, then it may be written as ${\bf T}(s)=(\cos\theta(s), \sin\theta(s))$, where $\theta(s)$ is another random variable.  In this way, the stochastic equations \eqref{ecsestom1} can be rewritten in the following manner
\begin{subequations}
\label{stochasticecs}
\begin{align}
\label{stochasticecs0}
\frac{d}{ds}{\bf R}\left(s\right)=&\left(\cos\theta(s),\sin\theta(s)\right),\\
\frac{d}{ds}\theta\left(s\right)=&\kappa(s).
\label{stochasticecs1}
\end{align}
\end{subequations}
The most important feature of these equations is their analogy with the Langevin equations for an active particle in the overdamped limit, where the noise is introduced through the stochastic curvature $\kappa(s)$ \cite{Fal-Sevilla2014}. Moreover, these equations can be studied through  traditional numerical methods, for example, using standard routines implemented in Brownian dynamics \cite{Fal-Ermak1978}.  Here, from an analytical viewpoint, we find it is more convenient to use a Fokker-Planck formalism in order to extract information of the above stochastic equations  (\ref{stochasticecs}).  

\subsection{From Frenet Stochastic Equations to the Hermans-Ullman Equation}

In this section, we present the Fokker-Planck formalism corresponding  to the stochastic equations \eqref{stochasticecs}. This description consists of determining the equation that governs the probability density function defined by
\begin{eqnarray}
P(\left.{\bf R}, \theta\right|{\bf R}^{\prime}, \theta^{\prime}; s)=\left<\delta({\bf R}-{\bf R}(s))\delta(\theta-\theta(s))\right>, 
\end{eqnarray}
where ${\bf R}$ and ${\bf R}^{\prime}$ are the ending positions of the polymer, and the angles $\theta$ and $\theta^{\prime}$ are their corresponding directions, respectively. The parameter $s$ is the polymer length.  

Applying the standard procedure described in Refs. \cite{Fal-Zinn1996, Fal-Gardiner1986} on the stochastic equations \eqref{stochasticecs}, we obtain the corresponding Fokker-Planck equation
\begin{eqnarray}
\frac{\partial P}{\partial s}+\nabla\cdot\left({\bf t}\left(\theta\right)P\right)=\frac{1}{2\ell_{p}}\frac{\partial^2P}{\partial\theta^2},
\label{F-P}
\end{eqnarray}
where ${\bf t}(\theta)=\left(\cos\theta,\sin\theta\right)$ and $\nabla$ is the gradient operator with respect to ${\bf R}$.  Let us look carefully at the last equation. Surprisingly, this equation is exactly the equation found by J. J. Hermans and R. Ullman in 1952 \cite{PSaito-Hermans1952}. They derived it supposing that the conformation of a polymer is determined by Markovian walks, taking the mean values of $\theta$ and $\theta^2$ as phenomenological parameters. These are parameters based on the X-ray dispersion experiments performed by Kratky and Porod \cite{PSaito-Kratky1949}.   For this reason, from now on, we name \eqref{F-P}  the Hermans-Ullman (HU) equation. It must be mentioned that H. E. Daniels found an equivalent equation few months before Hermans and Ullman \cite{PSaito-Daniels1952}. A revision of the methods used to obtain the HU equation can be found in Refs. \cite{Wlc-Yamakawa2016, Polcon-Chen2016}. For instance, taking into account \eqref{funcional}, HU can be derived through the Green formalism \cite{Polcon-Chen2016}. In contrast, in the present work, we have deduced the Hermans-Ullman equation considering two postulates, namely, I) the conformation of the semiflexible polymer satisfy the Frenet stochastic equations \eqref{ecsestom1}, and II) the stochastic curvature is distributed according to \eqref{dft}, which is consistent with the worm-like chain model (\ref{funcional}). As far as we know, this procedure has not been reported in the literature. 
 
 To end this section, let us remark, as it is pointed  out in   \cite{Wlc-Yamakawa2016, Polcon-Chen2016}, that
    \begin{eqnarray}
  \int d^{2}{\bf R}d^{2}{\bf R}^{\prime}~P(\left.{\bf R}, \theta\right|{\bf R}^{\prime}, \theta_{0}; s)\propto\mathbb{Z}\left(\theta,\theta_{0}, s\right),
  \end{eqnarray} 
 where $\mathbb{Z}\left(\theta,\theta_{0}, s\right)$ is  the marginal probability density function (see appendix \ref{A}), which establishes a bridge to the formalism in  N. Sait$\hat{\rm o}$ et al. \cite{PSaito-Saito1967} for the semiflexible polymer in the thermal bath.

\subsection{Multipolar decomposition for the Hermans-Ullman Equation} \label{sect3}
It is necessary to emphasize that the HU equation naturally arises in the description of the motion of an active particle. Thus, being careful with the right interpretation, the methods developed in Refs. \cite{Fal-Sevilla2014, Fal-Castro-Villarreal2018} to solve Eq.~\eqref{F-P} can be applied in this context.  Particularly, we use the multipolar expansion approach to solve Eq.~\eqref{F-P}, which in the orthonormal Cartesian basis $\{1, 2t_{i}, 4(t_{i}t_{j}-\frac{1}{2}\delta_{ij}), \cdots\}$, takes the following form  \cite{ Fal-Castro-Villarreal2018}
\begin{eqnarray}
P({\bf R},\theta, s)&=&\rho({\bf R}, s)+2\mathbb{P}_{i}({\bf R}, s) {t_{i}}\nonumber\\&+&4\mathbb{Q}_{ij}\left({\bf R},s\right)\left(t_{i}t_{j}-\frac{1}{2}\delta_{ij}\right)\nonumber\\
&+&8\mathbb{R}_{ijk}\left({\bf R}, s\right)\left(t_{i}t_{j}t_{k}-\frac{1}{4}\delta_{\left(ij\right.}t_{\left.k\right)}\right)+\cdots,\nonumber\\
\end{eqnarray}
where we have adopted the Einstein summation convention, and the symbol $(ijk)$ means symmetrization on the indices $i,j,k$. The coefficients of the series are multipolar tensors given by
 \begin{eqnarray}
\rho({\bf R}, s)&=&\int_{0}^{2\pi} \frac{d\theta}{2\pi}~P({\bf R},\theta, s)\nonumber,\\ 
\mathbb{P}_{i}({\bf R}, s)&=&\int_{0}^{2\pi} \frac{d\theta}{2\pi}~{t}_{i}~ P({\bf R}, \theta, s),\nonumber\\ 
\mathbb{Q}_{ij}({\bf R}, s)&=&\int_{0}^{2\pi} \frac{d\theta}{2\pi}~\left(t_{i}t_{j}-\frac{1}{2}\delta_{ij}\right)P({\bf R}, \theta, s), \nonumber\\
\mathbb{R}_{ijk}({\bf R}, s)&=&\int_{0}^{2\pi} \frac{d\theta}{2\pi}~\left(t_{i}t_{j}t_{k}-\frac{1}{4}\delta_{\left(ij\right.}t_{\left.k\right)}\right)P({\bf R}, \theta, s), \nonumber\\
&\vdots&.
\end{eqnarray}
In the latter coefficients, we have ignored the $\theta$ dependence of the vector ${\bf t}$ for reasons of notation. We also have obviated the dependence on ${\bf R}^{\prime}$ and $\theta^{\prime}$ to improve notation. The physical meaning of these tensors is as follows: $\rho({\bf R}, s)$ is the probability density function (PDF) of finding configurations with ends at ${\bf R}$ and ${\bf R}^{\prime}$, $\mathbb{P}({\bf R}, s)$ 
is the local average of the polymer conformational direction, $\mathbb{Q}_{ij}\left({\bf R}, s\right)$ 
is the correlation between the components $i$ and $j$ of the polymer direction ${\bf t}$, etc.

From Hermans-Ullman Eq.~\eqref{F-P}, it is possible to determine hierarchy equations for the multipolar tensors, which have already been shown for active particles in Refs. \cite{Fal-Sevilla2014, Fal-Castro-Villarreal2018}.
The same hierarchy equations can also be found in the semiflexible polymer context. Integrating over the angle $\theta$ in Eq.~\eqref{F-P}, we obtain the following continuity-type equation
\begin{eqnarray}
\frac{\partial \rho({\bf R}, s)}{\partial s}=-\partial_{i}\mathbb{P}^{i}\left({\bf R}, s\right). 
\label{eq1}
\end{eqnarray}
The related equation for $\mathbb{P}_{i}\left({\bf R},s\right)$ is obtained by multiplying Eq.~\eqref{F-P} by ${\bf t}(\theta)$, and using the definition of the tensor $\mathbb{Q}_{ij}({\bf R}, s)$. Thus, we found
\begin{eqnarray}
\frac{\partial \mathbb{P}_{i}({\bf R}, s)}{\partial s}=-\frac{1}{2\ell_{p}}\mathbb{P}_{i}({\bf R}, s)-\frac{1}{2}\partial_{i}\rho({\bf R},s)-\partial^{j}\mathbb{Q}_{ij}({\bf R}, s).\nonumber\\
\label{eq2}
\end{eqnarray}
In the same way, we obtain the equation for $\mathbb{Q}_{ij}({\bf R}, s)$,
\begin{eqnarray}
\frac{\partial\mathbb{Q}_{ij}({\bf R}, s)}{\partial s}=-\frac{2}{\ell_{p}}\mathbb{Q}_{ij}({\bf R}, s)-\frac{1}{4}\mathbb{T}_{ij}({\bf R}, s)-\partial^{k}\mathbb{R}_{ijk}({\bf R}, s), \nonumber\\
\end{eqnarray}
where $\mathbb{T}_{ij}$ denotes the second rank tensor $-\delta_{ij}\partial_{k}\mathbb{P}^{k}+\left(\partial^{i}\mathbb{P}^{j}+\partial^{j}\mathbb{P}^{k}\right)$. Similarly, the equations for the rest of tensorial fields can be computed recursively for consecutive ranks. Taking a combination of \eqref{eq1} and \eqref{eq2}, we observe that the PDF $\rho({\bf R}, s)$ and the two-rank tensor $\mathbb{Q}_{ij}({\bf R}, s)$ are involved in one equation given by
\begin{eqnarray}
\frac{\partial^2\rho({\bf R}, s)}{\partial s^2}+\frac{1}{2\ell_{p}}\frac{\partial\rho({\bf R}, s)}{\partial s}=\frac{1}{2}\nabla^2\rho({\bf R}, s)+\partial_{i}\partial_{j}Q^{ij}({\bf R}, s).\nonumber\\
\label{eq3}
\end{eqnarray}

It is noteworthy to mention, that Eq. (\ref{eq3}) is a modified version of Telegrapher's equation \cite{Fal-Masoliver1989}, where the term $\partial_{i}\partial_{j}Q^{ij}({\bf R}, s)$ makes the difference.  In the following, we use Eq. (\ref{eq3}) for the case of a semiflexible polymer in the open Euclidean plane as a test case. This allows us to verify the famous experimental result of Kratky and Porod \cite{PSaito-Kratky1949} using this procedure.

\subsubsection{Example: Testing the Kratky-Porod result.}\label{sectII}
In this section, we study the case of a semiflexible polymer on the Euclidean plane. In order to reproduce the well known result of Kratky-Porod \cite{PSaito-Kratky1949}, we apply the multipolar series method shown in the previous section to compute the mean square end-to-end distance.
The end-to-end distance is defined as $\delta{\bf R}:={\bf R}-{\bf R}^{\prime}$, thus the mean square end-to-end distance is given by 
\begin{equation}
 \left<\delta{\bf R}^2\right>\equiv\int_{\mathbb{R}^{2}\times\mathbb{R}^{2}}\rho({\bf R}|{\bf R}^{\prime}; s)\delta{\bf R}^2~d^{2}{\bf R}~d^2{\bf R}^{\prime}.
 \label{MS}
 \end{equation}
To compute this quantity, we use Eq.~\eqref{eq3} to show  that l.h.s of (\ref{MS}) satisfies 
  {\small\begin{eqnarray}
\frac{\partial^2\left<\delta{\bf R}^2\right>}{\partial s^2}+\frac{1}{2\ell_{p}}\frac{\partial\left<\delta{\bf R}^2\right>}{\partial s}&=&\int d^{2}{\bf R}~d^2{\bf R}^{\prime}\left(\delta{\bf R}\right)^2\times\nonumber\\&&\left[\frac{1}{2}\nabla^2\rho({\bf R}, s)+\partial_{i}\partial_{j}Q_{ij}({\bf R}, s)\right].\nonumber\\
\label{eq4}
\end{eqnarray}}
Integrating by parts on the r.h.s. of \eqref{eq4}  with respect to ${\bf R}$, using that $\nabla^2\delta{\bf R}^2=4$ and the traceless condition $\delta_{ij}\mathbb{Q}^{ij}=0$, we have that  $\left<\delta{\bf R}^2\right>$ satisfies the differential equation

\begin{eqnarray}
\frac{\partial^2\left<\delta{\bf R}^2\right>}{\partial s^2}+\frac{1}{2\ell_{p}}\frac{\partial\left<\delta{\bf R}^2\right>}{\partial s}=2.
\label{eq5}
\end{eqnarray}
Now, we solve this differential equation with the initial conditions, for $s=0$, $\left<\delta{\bf R}^2\right>=0$ and $\frac{d}{ds}\left<\delta{\bf R}^2\right>=0$. The final polymer length is denoted by $L$. 

In this way, we found that the mean square end-to-end distance is given by
\begin{eqnarray}
\left<\delta{\bf R}^2\right>=4\ell_{p}L-8\ell_{p}^2\left(1-\exp\left(-\frac{L}{2\ell_{p}}\right)\right),
\label{planoabierto}
\end{eqnarray}
which is the standard Kratky-Porod result for semiflexible polymers confined to a plane 
\cite{PSaito-Kratky1949, PSaito-Hermans1952}. The last result has two well-known asymptotic limits, namely, 
\begin{eqnarray}
\left<\delta{\bf R}^2\right>\simeq\left\{\begin{array}{cc}
 4\ell_{p}L, & {\rm if }~L\gg \ell_{p},\\
 &\label{asymptotics}\\ 
 L^2, & {\rm if}~L\ll\ell_{p}.\\
 \end{array}\right.
\end{eqnarray}

In the first case, the polymer conformations are equivalent to brownian trajectories. In this case, the polymer is called Gaussian polymer \cite{PDoi-Doi1988book}. In the second case, the polymer takes only one configuration; it goes in a straight line, which is known as the ballistic limit. We remark that the result in Eq.~\eqref{planoabierto} is usually obtained by using different analytical approaches (for example, see Appendix \ref{A} and Refs. \cite{MGD-Kamien2002, PSaito-Saito1967}).
  
  In the next section, we address the study of a confined polymer to a flat compact domain within the approach developed above.

\section{Semiflexible polymer in a compact plane domain }

\subsection{General expressions for a semiflexible polymer in an arbitrary compact domain}\label{sectTE}
In this section, we apply the hierarchy equations developed in section \ref{sect3} in order to determine the conformational states of a semiflexible polymer confined to a flat compact domain $\mathcal{D}$.  Commonly, it is necessary to truncate the hierarchy equations at some rank. For instance, at first order, let us consider  $\mathbb{P}_{i}({\bf R}, s)$ as a constant vector field, then (\ref{eq1}) implies  that  $\rho({\bf R}, s)$ is uniformly distributed, which clearly it is not an accurate description because otherwise it means that the mean square end-to-end distance would be a constant for all values of the polymer length $s$. An improved approximation consists of taking the truncation on the second hierarchy rank, which corresponds to assume that $\mathbb{Q}_{ij}({\bf R}, s)$ is uniformly distributed. Indeed, the truncation approximation gets better the larger the polymer length is, since as it is pointed out in \cite{Fal-Castro-Villarreal2018} from Eqs. (\ref{eq1}) and (\ref{eq2}) one can conclude that the tensors $\mathbb{P}_{i}({\bf R}, s)$ and $\mathbb{Q}_{ij}({\bf R}, s)$ damp out as $e^{-L/(2\ell_{p})}$ and $e^{-2 L/\ell_{p}}$, respectively. From these expressions, clearly $\mathbb{Q}_{ij}({\bf R}, s)$ damps out more strongly than $\mathbb{P}_{i}({\bf R}, s)$ for larger polymer length. In the polymer context, it means that the tangent directions of the polymer are uniformly correlated.

 In the following, let us define a characteristic length $a$ associated to the size of the compact domain $\mathcal{D}$, thus if we scale polymer length $s$ with $a$ one can consider $2a/\ell_{p}$ as a dimensionless attenuation coefficient associated to the damp out of $\mathbb{Q}_{ij}({\bf R}, s)$.  Thus as long as we consider cases when $2a/\ell_{p}$ far from 1, we may  neglect the contribution of $\mathbb{Q}_{ij}({\bf R}, s)$. Here, we are going to consider this latter case, therefore according to (\ref{eq3}), the Telegrapher's equation is the one considered as the governing equation of the PDF $\rho(\left.{\bf R}\right|{\bf R}^{\prime}, s)$, that is, 
\begin{eqnarray}
\frac{\partial^2\rho({\bf R}, s)}{\partial s^2}+\frac{1}{2\ell_{p}}\frac{\partial\rho({\bf R}, s)}{\partial s}=\frac{1}{2}\nabla^2\rho({\bf R}, s),
\label{eq6}
\end{eqnarray}
with the initial conditions 
\begin{subequations}
\label{IC}
\begin{align}
\label{ICa}
\lim_{s\to 0}\rho(\left.{\bf R}\right|{\bf R}^{\prime}, s)=&\delta^{\left(2\right)}({\bf R}-{\bf R}^{\prime}),\\
\lim_{s\to 0}\frac{\partial \rho(\left.{\bf R}\right|{\bf R}^{\prime}, s)}{\partial s}=&0.
\label{ICb}
\end{align}
\end{subequations}
These conditions have the following physical meaning. Clearly, Eq.~\eqref{ICa} means that the polymer ends coincide when the polymer length is zero, whereas Eq.~\eqref{ICb} means that the polymer length does not change spontaneously. Since the polymer is confined to a compact domain, we also impose a Neumann boundary condition
\begin{eqnarray}
\left.\nabla \rho\left(\left.{\bf R}\right|{\bf R}^{\prime}, s\right)\right|_{{\bf R}, {\bf R}^{\prime}\in\partial \mathcal{D}}&=&0, ~~~~\forall s, 
\label{BC}
\end{eqnarray}
where $\partial \mathcal{D}$ is the boundary of $\mathcal{D}$. 
This boundary condition means that the polymer does not cross the boundary coating the domain. 

To solve the differential equation \eqref{eq6}, we use the standard separation of variables \cite{Fal-Feshbach1953}.  This method requires to solve the so-called Neumann eigenvalue problem. It consists of finding all possible real values $\lambda$, for which there exists a non-trivial solution $\psi\in C^{2}(\mathcal{D})$ that satisfies the eigenvalue equation $-\nabla^2\psi=\lambda\psi$, and the Neumann boundary condition \eqref{BC}.  In this case, the set of eigenvalues is a sequence $\lambda_{{\bf k}}$ with ${\bf k}$ in a numerable set $I$, and each associated eigenspace is finite dimensional. These latter eigenespaces are orthogonal to each other in the space of square-integrable functions $L^2(\mathcal{D})$ \cite{Fal-Chavel1984, Fal-Feshbach1953}. That is, the sequence $\lambda_{\bf k}$ is associated with the set of eigenfunctions $\{ \psi_{\bf k}({\bf R})\}$ that satisfy the orthonormal relation
\begin{eqnarray}
\int_{\mathcal{D}}\psi_{\bf k}\left({\bf R}\right)\psi_{\bf k^{\prime}}\left({\bf R}\right)d^{2}{\bf R}=\delta_{{\bf k}, {\bf k}^{\prime}}. \label{ortogonal}
\end{eqnarray}

 Next, we expand the probability density function $\rho(\left.{\bf R}\right|{\bf R}^{\prime}, s)$ in a linear combination of those eigenfunctions $\{ \psi_{\bf k}({\bf R})\}$, that is, a spectral decomposition  $\rho(\left.{\bf R}\right|{\bf R}^{\prime}; s)=\sum_{{\bf k}}g_{\bf k}(s)\psi_{\bf k}({\bf R})\psi_{\bf k}({\bf R}^{\prime})$. Substituting this series in the Telegrapher's  equation \eqref{eq6}, we find that the functions $g_{\bf k}(s)$ satisfy the following ordinary differential equation
\begin{eqnarray}
\frac{d^2g_{\bf k}(s)}{ds^2}+\frac{1}{2\ell_{p}}\frac{dg_{\bf k}(s)}{ds}+\frac{1}{2}\lambda_{\bf k}g_{\bf k}(s)=0, 
\label{eq8}
\end{eqnarray}
where the initial conditions, (\ref{IC}), imply $g_{\bf k}(0)=1$, and $dg_{\bf k}(0)/ds=0$. Therefore, the solution is given by 
\begin{eqnarray}
g_{\bf k}(s)=G\left(\frac{s}{4\ell_{p}}, 8\ell_{p}^2\lambda_{\bf k}\right),
\end{eqnarray}
where
\begin{eqnarray}
{\scriptsize G(v, w)=e^{-v}\left[\cosh\left(v\sqrt{1-w}\right)+ \frac{\sinh\left(v\sqrt{1-w}\right)}{\sqrt{1-w}}\right]}.\nonumber\\
\label{Gfunction}
\end{eqnarray}

Finally, using the above information the probability density function is given by 
\begin{eqnarray}
\rho(\left.{\bf R}\right|{\bf R}^{\prime}, s)=\frac{1}{A\left(\mathcal{D}\right)}\sum_{{\bf k}\in I}G\left(\frac{s}{4\ell_{p}}, 8\ell_{p}^2\lambda_{\bf k}\right)\psi_{\bf k}({\bf R})\psi_{\bf k}({\bf R}^{\prime}), \nonumber\\\label{pdf}
\end{eqnarray}
where $A(\mathcal{D})$ is the area of the domain $\mathcal{D}$, which is needed in order to have a normalized probability density function in the space $\mathcal{D}\times{\mathcal{D}}$. Then, we have that $\rho(\left.{\bf R}\right|{\bf R}^{\prime}, s)d^{2}{\bf R}d^{2}{\bf R}^{\prime}$ is the probability of having a polymer in a conformational state with polymer length $s$, and ends at ${\bf R}$ and ${\bf R}^{\prime}$. Additionally, using the expression (\ref{pdf}), the mean square end-to-end distance can be computed in the standard fashion by
\begin{eqnarray}
\left<\delta{\bf R}^2\right>_{\mathcal{D}}=\sum_{{\bf k}\in I}a_{k}~G\left(\frac{s}{4\ell_{p}}, 8\ell_{p}^2\lambda_{\bf k}\right), 
\label{gen-sol}
\end{eqnarray}
where the coefficients $a_k$ are obtained by 
\begin{eqnarray}
a_{k}=\int_{\mathcal{D}\times{\mathcal{D}}}\left({\bf R}-{\bf R}^{\prime}\right)^2 \psi_{\bf k}({\bf R})\psi_{\bf k}({\bf R}^{\prime})~d^{2}{\bf R}~d^{2}{\bf R}^{\prime}.
\label{coef}
\end{eqnarray}

In the following, we shall discuss the specific case when the polymer is enclosed in a square box.

\subsection{Example: semiflexible polymer in a square domain}\label{thdomain}

In this section, we study the case when the semiflexible polymer is enclosed  in a square box $\mathcal{D}=\left[0,a\right]\times\left[0, a\right]$. For this domain, it is well known that the eigenfunctions of the Laplacian operator $\nabla^2$ correspond to a combination of products of trigonometric functions \cite{Fal-Chavel1984}. That is, for each pair of positive integer numbers $(n, m)$ and positions ${\bf R}=(x,y)\in\mathcal{D}$, it is not difficult to show  that the eigenfunctions of the Laplacian operator consistent with (\ref{BC}) are 
\begin{eqnarray}
\psi_{\bf k}\left({\bf R}\right)=\frac{2}{a}\cos\left(\frac{\pi n}{a}x\right)\cos\left(\frac{\pi m}{a}y\right).\nonumber
\end{eqnarray}
These functions constitute a complete orthonormal basis, that satisfy \eqref{ortogonal}. The corresponding eigenvalues are  $\lambda_{\bf k}={\bf k}^2$, with ${\bf k}=\left(\frac{\pi n}{a}, \frac{\pi m}{a}\right)$. 

Now, we proceed to determine the coefficients $a_{\bf k}$ \-using Eq.~\eqref{coef} in  order to give an expression for the mean square end-to-end distance. By straightforward calculation, the coefficients $a_{\bf k}$ are given explicitly by 
\begin{eqnarray}
a_{\bf k}=\left\{\begin{array}{cc}
\frac{1}{3}a^2,  & {\bf k}=0,\\
-\frac{4a^2}{\pi^4}\left(\frac{\left(1-(-1)^{n}\right)}{n^4}\delta_{m,0}+ \frac{\left(1-(-1)^{m}\right)}{m^4}\delta_{n,0}\right), & {\bf k}\neq 0.
\end{array}\right.\nonumber\\
\end{eqnarray}
Upon substituting the latter coefficients in the general expression \eqref{gen-sol}, we found that
\begin{eqnarray}
\frac{\left<\delta{\bf R}^2\right>_{\mathcal{D}}}{a^2}=\frac{1}{3}-\sum_{n\in 2\mathbb{N}+1 }\frac{32}{\pi^4 n^4}G\left(\frac{L}{4\ell_{p}}, 8\pi^2\left(\frac{\ell_{p}}{a}\right)^2n^2\right),\nonumber\\
\label{sol}
\end{eqnarray}
where $2\mathbb{N}+1$ is the set of odd natural numbers.  Since the function $G(v,w)$ satisfy that $G(v, w)\leq 1$ for all positive real numbers $v$ and $w$, the series in Eq.~\eqref{sol} is convergent for all values of $L/\ell_{p}$ y $\ell_{p}/a$. Considering this last property, it is possible to prove the following assertions.

\begin{prop} \label{result1}
Let $L/\ell_{p}$  any positive non-zero real number, then the mean square end-to-end distance \eqref{sol} obeys 
\begin{eqnarray}
 \lim_{\ell_{p}/a\to 0}\frac{\left<\delta{\bf R}^2\right>_{\mathcal{D}}}{\ell^2_{p}}=\frac{4L}{\ell_{p}}-8\left(1-\exp\left(-\frac{L}{2\ell_{p}}\right)\right). \nonumber
\end{eqnarray}
\end{prop}

\begin{prop}\label{result2} Let $L/\ell_{p}$ and $\ell_{p}/a$ any positive non-zero real numbers and $c=2/3-64/\pi^4$, then the mean square end-to-end distance \eqref{sol} obeys
\begin{eqnarray}
 0\leq \frac{\left<\delta{\bf R}^2\right>_{\mathcal{D}}}{a^2}\leq \frac{2}{3},\nonumber\end{eqnarray}
and
\begin{eqnarray}
 0\leq \frac{\left<\delta{\bf R}^2\right>_{\mathcal{D}}}{a^2}-\left(\frac{1}{3}-\frac{1}{3}G\left(\frac{L}{4\ell_{p}}, 8\pi^2\frac{\ell^2_{p}}{a^2}\right)\right)\leq c.\nonumber
\end{eqnarray}

\end{prop}

Claim {\bf \ref{result1}} recovers  the Kratky-Porod result about the mean square end-to-end distance (see Eq.~\eqref{planoabierto}). Claim {\bf\ref{result2}} means that the mean-square end-to-end distance is bounded from below by $0$ and  is bounded from above by $2/3 a^2$. In addition, this second claim also provides an approximation formula  for $\left<\delta{\bf R}^2\right>_{\mathcal{D}}$, that is,  for all values of $L/\ell_{p}$ and $\ell_{p}/a$ such that the condition
\begin{eqnarray}
1-G\left(\frac{L}{4\ell_{p}}, 8\pi^2\frac{\ell^2_{p}}{a^2}\right)\gg 3c\label{cond}
\end{eqnarray}
holds, one has the following approximation
{\small\begin{eqnarray}
\frac{\left<\delta{\bf R}^2\right>_{\mathcal{D}}}{a^2}&\simeq&\frac{1}{3}-\frac{1}{3}\exp\left(-\frac{L}{4\ell_{p}}\right)\nonumber\\
&\times&\left\{ \cosh\left[\frac{L}{4\ell_{p}}\left(1-8\pi^2\frac{\ell^2_{p}}{a^2}\right)^{\frac{1}{2}}\right]\right.\nonumber\\&+&\left.\left(1-8\pi^2\frac{\ell^2_{p}}{a^2}\right)^{-\frac{1}{2}}\sinh\left[\frac{L}{4\ell_{p}}\left(1-8\pi^2\frac{\ell^2_{p}}{a^2}\right)^{\frac{1}{2}}\right]\right\}.\nonumber\\ 
\label{approx2}
\end{eqnarray}}
Let us point out,  the validity of this approximation occurs provided that the condition (\ref{cond}) holds, that is, whenever one can neglect the value $c$ (See Appendix \ref{coefficients} for proofs of claims 1 and 2). 

In the following, let us remark that for any fixed value of $a$, the r.h.s. of (\ref{approx2}), as a function of $L$, shows the existence of a critical persistence length, $\ell^{*}_{p}\equiv a/(\pi\sqrt{8})$, such that for all values $\ell_{p}>\ell^{*}_{p}$ it exhibits an oscillating behavior, whereas for $\ell_{p}<\ell^{*}_{p}$, it is monotonically increasing. In addition, for each value of $\ell_{p}$ the function (\ref{approx2}) converges to $1/3$ as long as $L\gg a$. The critical persistence length, therefore,  distinguishes two conformational behaviors of the semiflexible polymer enclosed in the square box. 
In Fig. \ref{MSD-ex}, the mean-square end-to-end distance, Eq. (\ref{sol}) and r.h.s. of (\ref{approx2}),  have been shown  for the ratios $\ell_{p}/a=1/32, 1/16,1/8, 1/4,1/2, 1$ where we can appreciate both conformational states. Furthermore, one of the most intriguing features of the above approximation (\ref{approx2}) is its similar structure to the corresponding one in the case of a polymer wrapping a spherical surface. Indeed, let us remark that \eqref{approx2} has the same mathematical structure that the mean square end-to-end distance found by Spakowitz and Wang in \cite{PSaito-Spakowitz2003}, exhibiting both conformational states . 

In the next section, we address the study of the semiflexible polymer  through a Monte Carlo algorithm in order to corroborate the results found here.

\section{Monte Carlo-Metropolis algorithm for semiflexible polymers  }\label{mc}

Here, we develop a Monte Carlo algorithm to be use in computer simulations in order to study the conformational states of a semiflexible polymer enclosed in a compact domain. Particularly, the algorithm shall be suited to the square domain which, additionally,  will allow us to validate our analytical approximations shown in the latest section.
 
\begin{figure}[ht!]
\centering
\includegraphics[scale=.45]{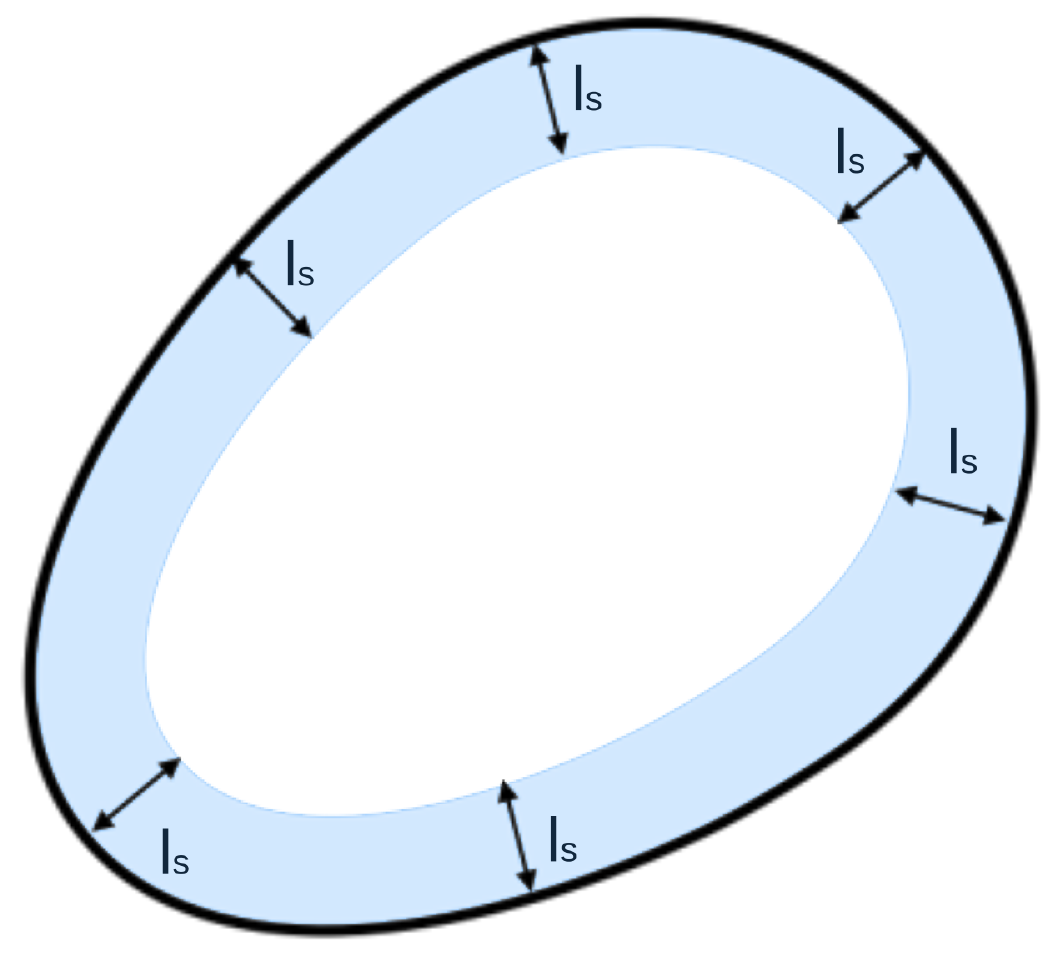}
\caption{Generic compact domain $\mathcal{D}$ is shown, where the boundary wall $\partial\mathcal{D}$ is represented by the black continuous line. The condition defining the beads near to the wall is represented by the blue filled region of width $l_s$. }
\label{frame}
\end{figure}

As we have emphasized above, the worm-like chain model is the suitable framework to describe the spatial distribution of semiflexible polymers, which are modeled as $n$ beads consecutively connected by $n-1$ rigid bonds, called Kuhn's segments \cite{Fal-Yamakawa1971}. Each bead works like a pivot allowing us to define an angle $\theta_i$ between two consecutive bonds, where $i$ is the label of the $i-$th bead. This model requires a potential energy description where all possible contributions due to bead-bead, bead-bond and bond-bond interactions are taken into account. In a general setting, energies of bond-stretching, elastic bond angle, electrostatic interaction, torsional potential, etc., should be considered, such as in Refs.~\cite{Wlc-Qiu2016, Psim-Chirico1994, Psim-Allison1989,Psim-DeVries2005,Psim-Jian1997,Psim-Jian1998}. However, here we are interested solely on the study of  possible spatial configurations of a single semiflexible polymer enclosed in a compact domain $\mathcal{D}$,  such as the one shown in Fig. \ref{frame}. Thus in our case we only take into account  two energetic contributions, namely, the elastic bond angle and the  wall-polymer interaction. The first contribution, that is, the elastic bond angle is given by
\begin{equation}
E_b=\frac{g}{2}\sum_i \theta_i^2,
\end{equation}
where $\theta_i$ is the angle between two consecutively bonds, and $g=\alpha/l_{0}$, where we recall that  $\alpha$ is the bending rigidity, and $l_{0}$ is the Kuhn length. In addition, we must consider the  wall-polymer interaction given by
\begin{eqnarray}
  E_{w}=
  \begin{cases}
                                   0, & \text{if all beads are in $\mathcal{D}$}, \\
                                   \infty, & \text{if there are beads outside of $\mathcal{D}$}.
  \end{cases}
\end{eqnarray}

In the algorithm, the acceptance changes criteria of the polymer spatial configurations take the structure of a Gaussian distribution function. In this context, we generate random chains enclosed in $\mathcal{D}$, constituted by $N$ bonds with constant Kuhn length, implementing a growth algorithm. Our computational realization consists of bead generation attending the following conditions:
{\it starting bead}, {\it beads far from walls}, {\it beads near to walls}, and {\it selection problem}
which are explained in the following subsections. 

\subsection*{Starting bead}

This condition  describes the process of initial bead generation and the acceptance criteria of the second bead. Let us regard, without loss of generality that the origin of $\mathbb{R}^2$ belongs to $\mathcal{D}$. We choose the initial bead  at $\mathbf{x}_0$ as a uniformly distributed random point in the region $\mathcal{D}$. We define the auxiliary vector $\mathbf{R}_{l_0}=(l_0,0)$, which is parallel to the horizontal axis, it will allow us to determine if the next bead is inside $\mathcal{D}$. Also, we consider the angle $\theta_0$ as the one that is formed between the horizontal axis and the first bead, which is taken from a uniform distribution in the interval $[0,2\pi]$. Now, we compute the following vector
\begin{equation}
\mathbf{R}'=\mathcal{R}
\left(\theta_0\right){\mathbf{R}_{l_0}}^T,
\end{equation}
where $\mathcal{R}
\left(\theta_0\right)$
denotes the two-dimensional rotational matrix by an angle $\theta_0$, defined as
\begin{equation}
\mathcal{R}\left(\theta_0\right)=
\begin{bmatrix}
\cos \theta_0 & -\sin \theta_0 \\
\sin \theta_0 & \cos \theta_0 \\
\end{bmatrix},
\end{equation}
and the superscript denotes the transposition of $\mathbf{R}_{l_0}$. The resultant vector $\mathbf{x}_0+\mathbf{R}'^T$ will be the position of the second bead only if this is inside $\mathcal{D}$. If this happened,  the new bead will be denoted by $\mathbf{x}_1$ as well as the vector $\mathbf{R}_{l_0}=\mathbf{R}'^T$ is actualized. On the contrary, we repeat the process until finding an angle that satisfies the condition that the second bead is enclosed in the domain.

\subsection*{Beads far from walls}\label{bffw}

This condition describes the method of subsequence bead generation. We say that a bead is far from walls if the perpendicular distance between the boundary $\partial\mathcal{D}$ and the bead is greater than a particular distance $l_{s}$. In this case, if the $(k-1)$-th bead satisfies this condition, we generate the subsequent bead taking a random angle $\theta_{k}$ distributed according to a Gaussian density function
 $\mathcal{N}(0,l_{0}/\ell_{p})$, where we recall $\ell_{p}$ as the polymer persistence length. As in the previous condition, we compute the vector $\mathbf{R}'=\mathcal{R}\left(\theta_0\right){\mathbf{R}_{l_0}}^T$ corresponding to the $k$-th rotation of $\mathbf{R}_{l_0}$. If $l_0<l_{s}$,
 the resultant vector $\mathbf{x}_{k-1}+\mathbf{R}'^T$ is in $\mathcal{D}$ for any angle. Therefore, all rotations are accepted, so we assign the position $\mathbf{x}_{k}$ to the $k$-th bead. 

\subsection*{Beads near walls}

\begin{figure}[ht!]
\centering
\includegraphics[scale=.6]{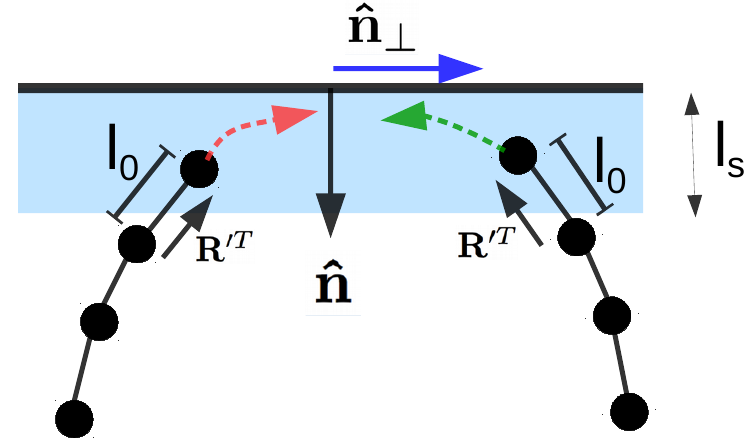}
\caption{Local schematic representation of the bending conditions when the beads are near to walls and $\mathbf{\hat{n}}\cdot\mathbf{R'}^T <0$.}
\label{bend}
\end{figure}

This condition describes the interaction between the polymer and the walls. The most important problem to solve here is the smooth bending of the chain when the polymer is near the walls. In a similar fashion as in the previous condition, we say that a bead is near the walls if the perpendicular distance from $\partial\mathcal{D}$ to the bead is smaller than $l_{s}$.  This condition looks like a frame surrounding the boundary of $\mathcal{D}$ (see blue filled region in Fig \ref{frame}). The scalar product $\mathbf{\hat{n}}\cdot\mathbf{R'}^T $  is used to seek the generation of a new bead, where $\mathbf{\hat{n}}$ is the normal vector of $\partial\mathcal{D}$, and $\mathbf{R'}^{T}$ is the orientation of the bead with position $\mathbf{x}_{k}$. In this region, we favor the smooth bending of the polymer taking into account the following rules. If $\mathbf{\hat{n}}\cdot\mathbf{R'}^T \geq0$, the generation of the new bead is going away from the wall. In this case, we generate the next bead according to second condition. If $\mathbf{\hat{n}}\cdot\mathbf{R'}^T <0$, the new bead generated approaches the wall. In this case, we promote the bending of the chain generating angles with the same Gaussian function as in the previous condition. However, we do the rotation by an angle $-\text{sgn}(R'^T_2)|\theta_k|$, where $\text{sgn}$ is the sign function, and $R'^T_2$ is the projection of $\mathbf{R'}^T$ over the perpendicular vector to $\mathbf{\hat{n}}$ (a $\pi/2$ rotation of $\mathbf{\hat{n}}$). The sign ``$-$'' appears there because we do counterclockwise rotations. Also, we need to check that the resultant vector $\mathbf{x}_{k-1}+\mathbf{R}'^T$ is  in $\mathcal{D}$, if this is the case, we assign the position $\mathbf{x}_k$ to the $k-$th bead. Otherwise, we generate angles $\theta_{ki}$ with a Gaussian distribution function $\mathcal{N}(0,l_0/\ell_p)$ until the next bead be in $\mathcal{D}$, doing rotation by angles
\begin{equation}
\theta_k=-\text{sgn}(R'^T_2)\sum_{i=1}^m|\theta_{ki}|,
\end{equation}
where $m$ is the number of angles generated until the next bead is in $\mathcal{D}$. In Fig.~\ref{bend}, we show a schematic representation of this conditions. Dashed arrows (green and red for positive and negative rotations, respectively) are the promoted direction according to the sign of the projection of $\mathbf{R}'^T$ over the perpendicular vector to  $\mathbf{\hat{n}}$ (blue arrow, denoted by $\mathbf{\hat{n}}_\bot$).

Finally, these steps are repeated $n-$times to generate a polymer of $n$ bonds (or $n+1$ beads) taking care of polymer-walls interaction. This algorithm allows us to generate the spatial configurations for confined polymers into a compact domain. It is clear that the total length of a polymer of $n$ bonds is $L=nl_0$. We denote the $\mathbf{x}(L)$ of the last bead after $n$ bonds so the mean square end-to-end distance is computed as $\left<{\delta \bf R}^2\right>_{\mathcal{D}}=\langle (\mathbf{x}(L)-\mathbf{x}_0)^2 \rangle$.  In Sec.~\ref{results} we shall analyze the results obtained for $\left<{\delta \bf R}^2\right>_{\mathcal{D}}$ using this algorithm as a function of the persistence length and the polymer length when $\mathcal{D}$ is a square box.

\subsection*{Selection problem }

The selection problem consist of choosing the adequate value of $l_{s}$. This value should be suitable to avoid the over or under bending of the polymer. For instance, if $\ell_p$ is comparable with the size of $\mathcal{D}$, and $l_s$ is not appropriate to promote the chain bending when the polymer is near the boundary $\partial\mathcal{D}$, the generation of beads outside of the domain will be favorable.  Therefore, the polymer will present bendings with high values of angles where the chain meets the boundary. The selection problem is resolved using the dimensional analysis of $\left<\delta{\bf R}^2\right>_{\mathcal{D}}$. Indeed, observe that in the continous limit of the chain, the mean-square end-to-end distance can be written as $\left<\delta{\bf R}^2\right>_{\mathcal{D}}=a^2 g(\ell_{p}/a, L/a)$, where $g(\ell_{p}/a, L/a)$ is a dimensionless function. Then, we choose $l_{s}$ such that the mean-square end-to-end distance computed with the simulation data depends just on this combination $\ell_{p}/a$ and $L/a$. In other words,  for $k$ we calculate $k$ profiles of $\left<\delta{\bf R}^2\right>_{\mathcal{D}}/a^2$ for a $k$ number of pairs $(\ell^{1}_{p}, a^{1}),(\ell^{2}_{p}, a^{2}), \cdots, (\ell^{k}_{p}, a^{k})$, with $\ell^{i}_{p}/a^{i}$  fixed for all $i=1,\cdots, k$, then we choose $l_{s}$ such that all these profiles collapse in a unique curve. 

\section{Semiflexible polymers enclosed in a square box: simulation vs analytical results}\label{results}
In this section, we are going to implement the algorithm explained in the preceding section for the particular case of a polymer enclosed in a square box of side  $a$. In this case, let us first note that for beads near the corners, checking the conditions to promote the chain bending for both adjacent walls at the same time is needed. 
Next, in the simulation we set up our unit length  by $d=10^{2}~l_{0}$. Now, we have to present the selection of $l_{s}$ according to the last part of the general algorithm. Thus, for the fixed ratios $\ell_{p}/a=1/50, 1/32, 1/16,1/8, 1/4,1/2, 1$, we study three profiles corresponding to the values $a/d=5,10,15$, respectively. In Table \ref{tab1} the selection values $l_{s}$ are shown once we collapse the three profiles in a unique curve. 

\begin{table}[h!]
\caption{\label{tab1}Values of $l_s$ used in simulations for different values of persistence length $\ell_p$.}
\begin{tabular}{|c| c|}
\hline
$\ell_p/a$  & $ l_s/a$\\
\hline
1 & 0.085\\
1/2 &  0.065\\
1/4 &  0.050\\
1/8 &  0.040\\
1/16 &  0.010\\
1/32 &  0.005\\
$\leq$1/50 & 0 \\
\hline
\end{tabular}
\end{table}

The results shown in this section were computed as the average over $10^6$ spatial configuration of confined polymers, which were obtained using the algorithm described in the previous section. In particular, we shall study two respective regimes defined through the comparison between the box side $a$ and the polymer length $L$. The first one, polymers in weak confinement whenever $L/a\leq 1$ is discussed in subsection~\ref{noconfinado}. The second one,  polymers in strong confinement, corresponding to the situation when the polymer length is larger than the box side, is  discussed in subsection ~\ref{confinado}.

\subsection{Polymer in weak confinement}\label{noconfinado}

\begin{figure}[ht!]
\centering
\includegraphics[scale=1.]{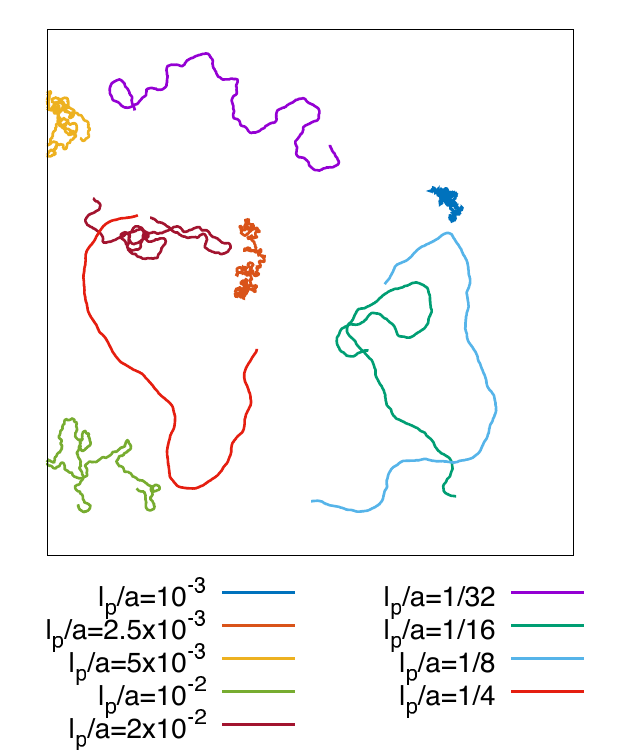}
\caption{Examples of semiflexible polymers, with length $L=a$, in the weak confinement regime for several values of persistence length. Solid black lines represent the walls of the box.}
\label{ex-short}
\end{figure}

In this regime, once we have solved the selection problem of $l_{s}$ we present the simulation results for polymers enclosed in a square box of side $a=10~d$ for different values of persistence lengths. In Fig.~\ref{ex-short}, examples of semiflexible polymers in weak confinement are shown. Notice that for very short persistence length, $\ell_p\simeq l_0$, the chain looks like a very curly string  like random walks, whereas when $\ell_p$ increases, the polymer adopts uncoiled configurations.

\begin{figure}[ht!]
\centering
\includegraphics[scale=1]{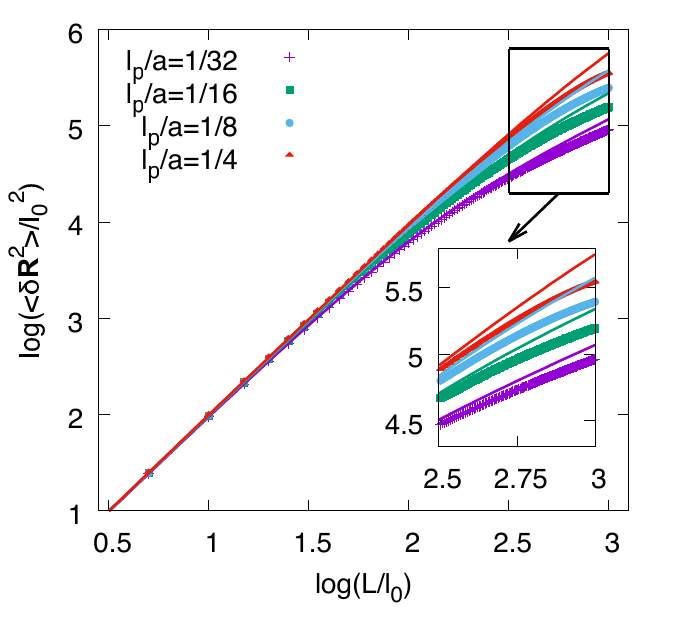}
\caption{Mean square end-to-end distance for confined polymers (point) with several persistence lengths. The deviation from the predictions for a polymer in a infinite plane (solid lines) given by Eq.~\eqref{planoabierto} is because on average the chain meets the polymer at lengths $L\simeq a$.}
\label{fig-MSDs}
\end{figure}

\begin{figure}[ht!]
\centering
\includegraphics[scale=1.]{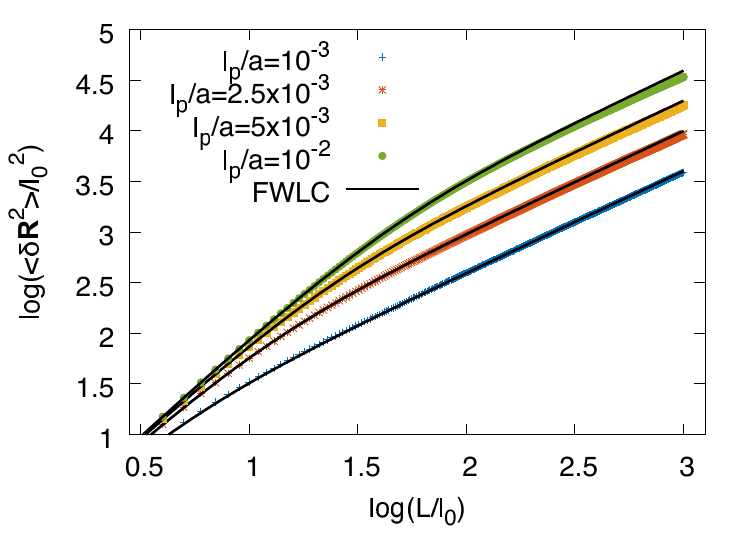}
\caption{Mean square end-to-end distance for polymers in the gaussian chain limit generated by the algorithm described in Sec.~\ref{mc}.}
\label{gchain}
\end{figure}

\begin{figure}[ht!]
\centering
\includegraphics[scale=1.]{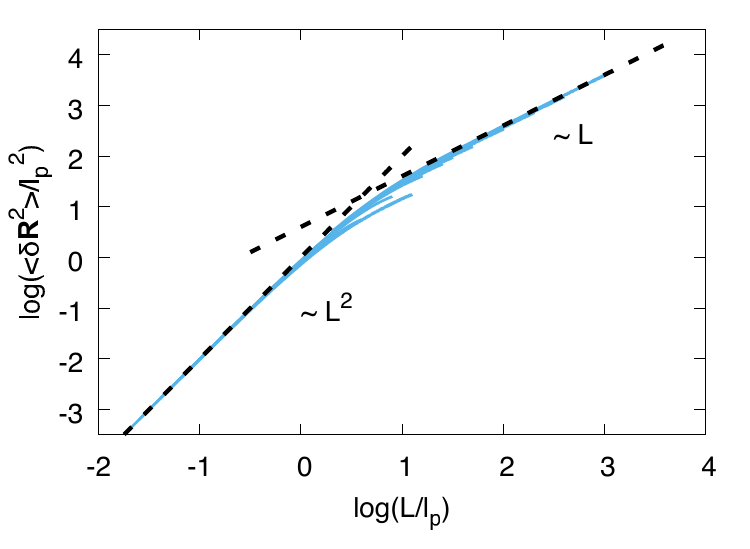}
\caption{Universal behavior of the Mean Square end-to-end distance in Fig.~\ref{fig-MSDs} in the scaling $\langle {\delta \bf R}^2 \rangle/ l_p^2$ vs $L/l_p$ (solid lines). Dashed lines are references for the ballistic and diffusive regimes for polymers in weak confinement.}
\label{fig-MSDs-Wang}
\end{figure}

The mean-square end-to-end distance is shown in Fig. \ref{fig-MSDs}, where  it is noted that it increases as long as $\ell_p$ increases, allowing the polymer to explore more surface as a function of the total polymer length. Also, notice that for small polymer lengths, the mean square end-to-end distance is in an excellent agreement with the results for semiflexible polymers in an infinite plane given by Eq.~\eqref{planoabierto}. Conversely, for polymer lengths around $L\simeq a$ we observe an slight deviation between the mean square end-to-end distance and the infinite plane solution (\ref{planoabierto}), because of the finite-size of the box. Notwithstanding, for small persistence lengths ($\ell_p\simeq 10^{-3}a$),  the mean square end-to-end distance is well fitted by Eq.~\eqref{planoabierto}. In this case, the polymer does not seem to be affected by the walls, since the  area explored by the chain is too short, on average, to meet the walls.

Furthermore, when the mean square end-to-end distance and the polymer length is scaled by $l_p^2$ and $l_p$, respectively, we found that the data shown in Fig.~\ref{fig-MSDs} and Fig.~\ref{gchain} are collapsed, respectively, into an unique plot shown in Fig.~\ref{fig-MSDs-Wang}, evidencing a ballistic behavior for small values of $L/\ell_p$ followed by a ``diffusive'' regime for large values of $L/\ell_{p}$. These results correspond to the well-known asymptotic limits of the Kratky-Porod result (\ref{asymptotics}). In addition, it is noteworthy to mention that these asymptotic limits are also  reported in \cite{PSaito-Spakowitz2003} for a semiflexible polymer wrapping a spherical surface in the corresponding plane limit.  

\subsection{Polymer in strong confinement}\label{confinado}

\begin{figure}[h!]
\centering
\includegraphics[scale=1]{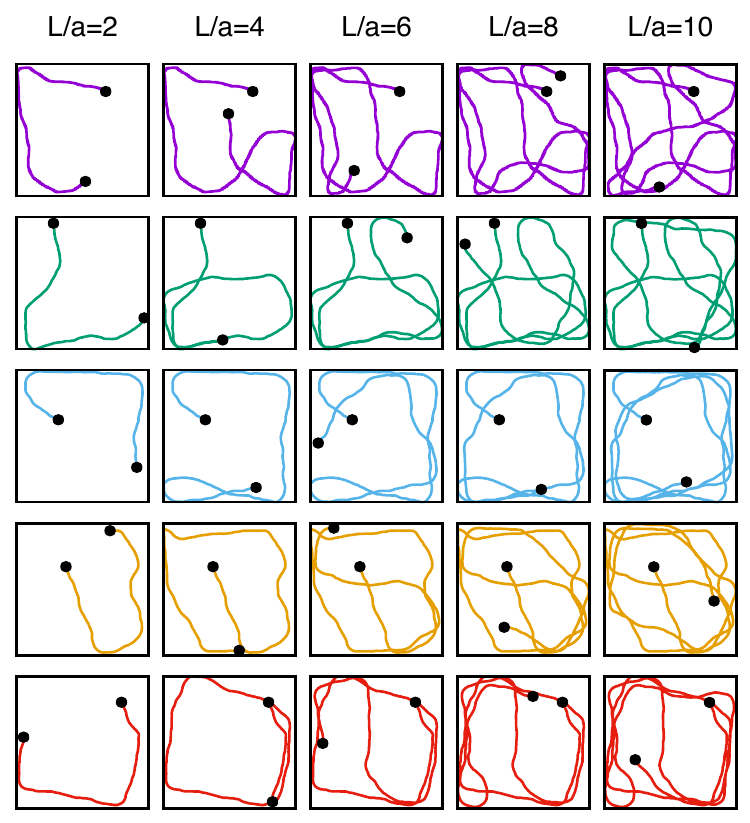}
\caption{Semiflexible polymer realizations for the value of persistence length $l_p=a$. Each column shows a polymer with a particular  length from the left to the right  with L/a=2, L/a=4, L/a=6, L/a=8, and L/a=10, respectively. Black filled circles indicate the endings of the polymer. It is also shown how the polymer rolling up around the square box while the polymer length becomes larger. }
\label{evo-ex}
\end{figure}

\begin{figure*}
\centering
\includegraphics[scale=.68]{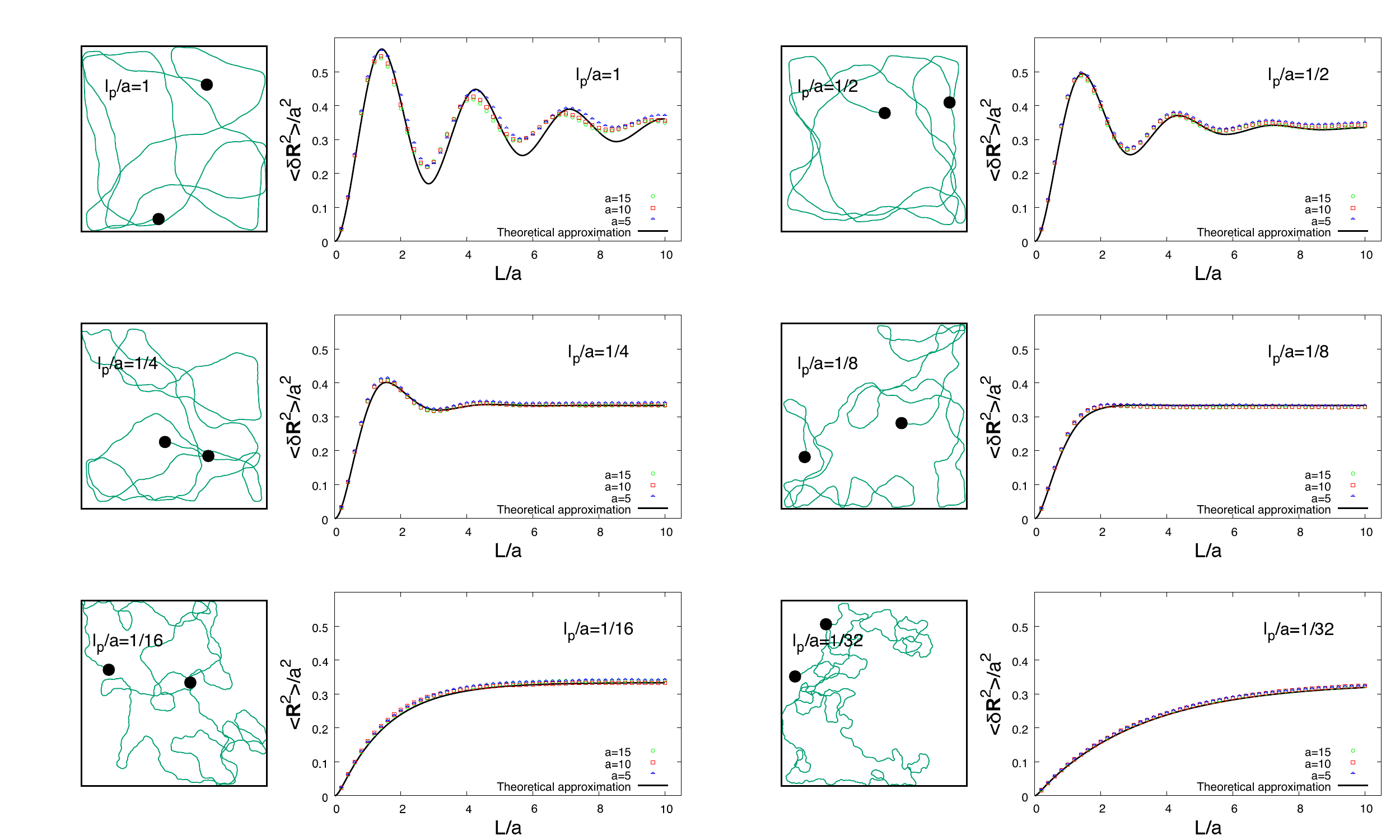}
\caption{\small Examples of polymers, in the first and third columns, are shown as solid green lines, where the initial and final beads are represented by black filled circles, and solid black lines represent the walls of the box. The figure also shows, in the second and fourth columns, the mean-square end-to-end distance for polymers in strong confinement as a function of the polymer length: The bold black line represents the superposition of the theoretical prediction (Eqs. (\ref{sol}) and (\ref{approx2})) with the simulation results shown for different box sides $a/d=5$ (blue triangles), $a/d=10$ (red squares) and $a/d=15$ (green circles). }
\label{MSD-ex}
\end{figure*}

In this section, we discuss the case of a polymer enclosed in a square box when its length is large enough to touch the walls, and to interact several times with them.  We perform simulations in order to generate polymers of lengths up to $L/a=10$ (chains of $10^4$ beads) for persistence lengths $\ell_{p}/a=1/32, 1/16,1/8, 1/4,1/2, 1$ for the values of  box side $a/d=5,10,$ and $15$, respectively. These simulations use the values of $l_s$ shown in Table~\ref{tab1}. Examples of these polymers are shown in  Fig. \ref{evo-ex} and in Fig.~\ref{MSD-ex}. 
\begin{figure}[ht!]
\centering
\includegraphics[scale=1]{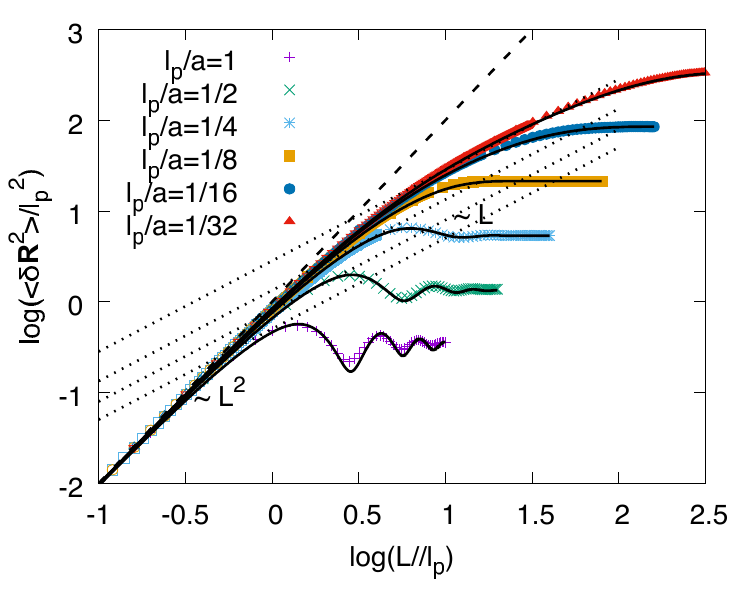}
\caption{\small Mean square end-to-end distance for polymers in strong confinement as a function of the persistence length. Note that $\langle \delta \mathbf{R} ^2 \rangle$ shows an oscillating behavior for values of persistence length satisfying the relation $\ell_p/a>8$, which is the same signature for the mean square end-to-end distance for polymers confined into a sphere. Dashed and doted lines has been plotted as references for the ballistic and diffusive behaviors, respectively.}
\label{MSD-Wang-strong}
\end{figure}
\begin{figure}[ht!]
\centering
\includegraphics[scale=1]{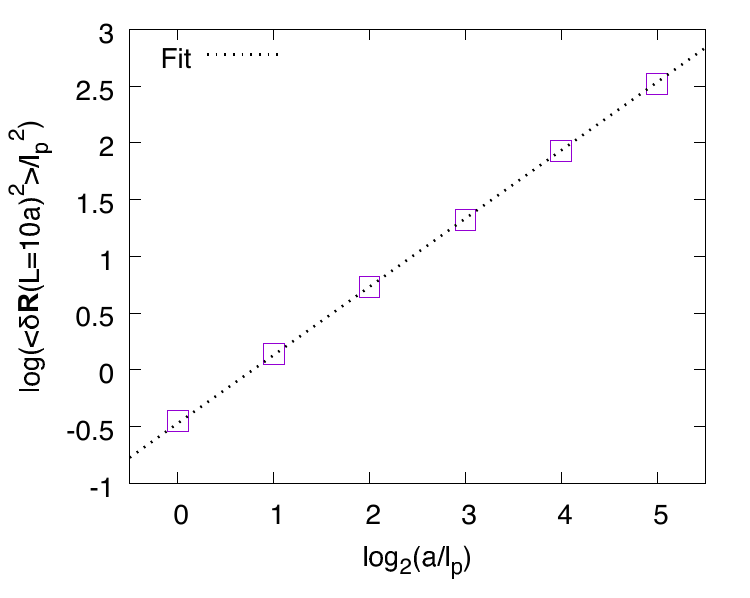}
\caption{\small Convergence of the mean square end-to-end distance $\langle \delta \mathbf{R}^2\rangle_{\mathcal{D}}=a^2/3$ for very large polymers in strong confinement as a function of the ratio $a/l_p$.}
\label{MSD-conv}
\end{figure}

In Fig.~\ref{MSD-ex}, we also report the mean square end-to-end distance scaled by $a^2$ as a function of $L/a$ for the different box sides ($a=$ 5, 10 and 15). In addition,  in Fig. \ref{MSD-Wang-strong} we report as well the mean square end-to-end distance scaled by $\ell_{p}^2$ as a function of $L/\ell_{p}$.   By simple inspection, in both cases an oscillating behavior is exhibited for values $\ell_p/a>\frac{1}{8}$, whereas a monotonically increasing behavior becomes evident for persistence lengths such that $\ell_{p}/a<\frac{1}{8}$. A growth in the number of oscillations is observed while $\ell_{p}/a$ increases from $1/8$. In addition, for values $\ell_{p}/a$ less than $1/8$ the behavior of the mean-square end-to-end distance corresponds to the one of a Gaussian polymer enclosed in a box and the corresponding conformational realization of the polymer looks like a confined random walk. As was mentioned before this transition between the oscillating and monotonic behavior of the conformation of the semiflexible polymer is very similar to that described by Wang and Spakowitz in Ref.~\cite{PSaito-Spakowitz2003} for semiflexible polymer confined to a spherical surface. Moreover, as noted in Fig. \ref{MSD-Wang-strong}  in the small polymer lengths regime, the mean square end-to-end distance shows a ballistic behavior, followed by a brief interval with a ``diffusive'' behavior.  Furthermore, an interesting observation is that the mean square end-to-end distance exhibits asymptotic plateau behavior for large values of $L/l_p$ as a function of $a/l_p$. In Fig.~\ref{MSD-conv} we show the value of the mean square end-to-end distance for the the polymer length $L=10~a$ in logarithmic scale as a function of $a/l_p$ in binary logarithmic scale. In this conditions, the plateau behavior is well fitted by a linear function, where the slope, $m$, of the line satisfies $m/\log2=1.97\sim 2$, whereas that the intercept takes the value $b=-0.473\pm0.006$. This last fact leads us to a universal scaling law of the mean square end-to-end distance regarding the box side for very large polymers:
\begin{equation}
\frac{\langle \delta \mathbf{R}(L=10a)^2\rangle}{a^2}=10^b\sim 0.336\pm 0.004,
\label{eq-msd-a}
\end{equation}
where the error has been computed as the  propagation error from the linear fit of the data in Fig.~\ref{MSD-conv}. This result is the universal convergence of the rate $\langle \delta \mathbf{R}^2\rangle_{\mathcal{D}}/a^2$ to $1/3$ for very large polymers, which becomes independent of the box side when the quotient $l_p/a$ keeps fixed. This is comprised, if we consider all available space in the box occupied  so that  a uniform distribution of beads occurs. Indeed, through the definition (\ref{MS}) with $\rho(\left. {\bf R}\right|{\bf R}^{\prime}; L\to\infty)=1/a^4$ one can get the desired result. Finally, it is noticeable that by simple inspection of the five polymer realizations shown in Fig. \ref{evo-ex}, for  $\ell_{p}/a=1$, all have the same conspicuous relation between the period of oscillation and the number of turns that the polymer performs. 

All these features of the behavior of the mean-square end-to-end distance are reproduced by the theoretical prediction (\ref{approx2}) and (\ref{sol}). In particular, it is significant to note that the critical persistence length found in our earlier discussion (see section (\ref{thdomain})) satisfies approximately $\ell^{*}_{p}/a=1/(\pi\sqrt{8})\approx 1/8$. We also note that for $\ell_{p}/a=1$ there is a slight discrepancy between the simulations results and the theoretical prediction (\ref{sol}) appearing in the three local minima shown in Fig. \ref{MSD-ex}. This is  due to the fact that for values of $\ell_{p}/a$ near $2$ there is a breakdown of the Telegrapher approximation that we performed in the section \ref{sectTE}.  In other words, the small disagreement appears for $\ell_{p}/a\approx 1$ because the role of the tensor $\mathbb{Q}_{ij}$ becomes important and it can not be neglected in Eq. (\ref{eq3}).

\section{Concluding remarks and perspectives}\label{conclusions}

In this paper, we have analyzed the conformational states of a semiflexible polymer enclosed in a compact domain. The approach followed rests on two postulates, namely, that the conformation of a semiflexible polymer satisfies the Frenet stochastic equations \eqref{ecsestom1}, and the stochastic curvature is distributed according to \eqref{dft}, which is consistent with the worm-like chain model. In addition, it turned out that the Fokker-Planck equation, corresponding to the stochastic Frenet equations,  is exactly the same as the Hermans-Ullman equation (see Eq. (\ref{F-P})) \cite{PSaito-Hermans1952}. Furthermore, taking advantage of the analogy between the Hermans-Ullman equation and the Fokker-Planck equation for a free active particle motion \cite{Fal-Castro-Villarreal2018}, we establish a multipolar decomposition for the probability density function, $P(\left.{\bf R}, \theta\right|{\bf R}^{\prime}, \theta^{\prime}; L)$, that describes  the manner in which a polymer with length $L$ distributes in the domain with certain endings, ${\bf R}$ and ${\bf R}^{\prime}$, and their associated directions $\theta$ and $\theta^{\prime}$, respectively.  In consequence, exploiting this analogy we provide an approximation for the positional distribution $\rho({\bf R}, {\bf R}^{\prime}, L)$ through Telegrapher's Equation, which for a compact domain is a good approximation as long as $2a/\ell_{p}>1$, where $a$ is a characteristic length of the compact domain.  In particular, we derive results for a semiflexible polymer enclosed in a square box domain, where we can give a mathematical formula for the {\it mean-square end-to-end distance}.  

Furthermore, we have developed a Monte Carlo-Metropolis algorithm to study the conformational states of a semiflexible polymer enclosed in a compact domain. In particular, for the square box domain, we compare the results of the simulation with the theoretical predictions finding an excellent agreement. Particularly, we have considered two situations, namely, a {\it polymer in weak confinement} and a {\it polymer in strong confinement} corresponding to polymers with length lesser and greater than the box side, respectively. In the weak confinement case, we reproduce the two-dimensional solution of a free chain i.e. the  Kratky-Porod result for polymers confined in two-dimensions. In the strong confinement case, we showed the existence of a critical persistent length $\ell^{*}_{p}\simeq a/8$ such that for all values $\ell_{p}>\ell^{*}_{p}$ the mean-square end-to-end distance exhibits an oscillating behavior, whereas for $\ell_{p}<\ell^{*}_{p}$, it is monotonically increasing. In addition, for each value of $\ell_{p}$ the function converges to $1/3$ as long as $L\gg a$. The critical persistence length, thus,  distinguishes two conformational behaviors of the semiflexible polymer enclosed in the square box. As was mentioned above, this result is the same type to the one found by Wang and Spakowitz in \cite{PSaito-Spakowitz2003} for a semiflexible polymer wrapping a spherical surface.  As a consequence of this resemblance, one can conclude that the shape transition from oscillating to monotonic conformational states provides evidence of a universal signature for a semiflexible polymer enclosed in a compact space. 

Our approach can be extended in various directions. For instance, the whole formulation can be extended easily to semiflexible polymers in three dimensions.  Although,  we must consider an stochastic version of the Frenet-Serret equations, now in this case, one would obtain the three dimensional case of the Hermans-Ullman equation, because the worm-like chain model involves just the curvature. In addition, the approach developed here can also be extended to the case where the semiflexible polymer wraps a curved surface.  

\section*{Acknowledgement}
P.C.V. acknowledges financial support by CONACyT Grant No. 237425 and PROFOCIE-UNACH 2017. J.E.R. acknowledges financial support from VIEP-BUAP (grant no. VIEP2017-123). The computer simulations were performed at the LARCAD-UNACH.

\appendix

\section{Sait\^{o}'s et al. Approach}\label{A}

The conformational states space of the polymer in this approach corresponds to the functional space $\{{\bf T}\left(s\right)|s\in\left[0, L\right]\}$, where ${\bf T}(s)$ is a tangent vector to the curve that describes the conformation of the polymer. 
Thus, the probability of having the polymer in a particular conformation given the ends directions at ${\bf T}_{L}$ and ${\bf T}_{0}$, is denoted by $P\left[{\bf T}\right]\mathcal{D}{\bf T}$. Upon integrating over all conformations with the exception of the fixed ends in  ${\bf T}_{L}$ and ${\bf T}_{0}$, we found that
\begin{eqnarray}
\int_{{\bf T}_{0}}^{{\bf T}_{L}}P\left[{\bf T}\right]\mathcal{D}{\bf T }=\frac{1}{\mathcal{N}}\mathbb{Z}\left({\bf T}_{L}, {\bf T}_{0}, L\right), 
\end{eqnarray}
where $\mathcal{D}{\bf T}$ is a functional measure, and $\mathbb{Z}$ is usually called the partition function \cite{MGD-Kamien2002}.
Note that $\mathbb{Z}\left({\bf T}(L), {\bf T}_{0}, L\right)/\mathcal{N}=P\left({\bf T}_{L}, {\bf T}_{0}, L\right)$ is the probability density of finding the polymer with ends directions ${\bf T}_{0}$ and ${\bf T}_{L}$, and length $L$. Also,
\begin{equation}
 \mathcal{N}=\int_{{\bf T}^{2}=1} d^{2}{\bf T}~\mathbb{Z}\left({\bf T}, {\bf T}_{0}, L\right),
 \end{equation}
 is the normalization constant.  
 The partition function is a path integral in the context of the path integral formulation in quantum mechanics developed by R. Feynman. Particularly, the partition function for the semiflexible polymer conformation in Sait\^{o}'s et al. approach is given by
\begin{equation}
\mathbb{Z}\left({\bf T}(L), {\bf T}_{0}, L\right)=\int_{{\bf T}\left(0\right)}^{{\bf T}\left(L\right)} \mathcal{D}{\bf T}\exp\left(-\frac{\ell_{p}}{2}\int_{0}^{L} \kappa^2 ds\right),
\label{Feynman}
\end{equation}
where  $\mathcal{D}{\bf T}$ is the functional measure for the polymer conformations. The partition function $\mathbb{Z}\left({\bf T}(L), {\bf T}_{0}, L\right)$  can be determined in several ways \cite{Fal-Zinn1996}. 

In this description, the average value of physical observables are computed in the traditional way
\begin{eqnarray}
\left<\mathcal{O}\left({\bf T}\right)\right>=\int_{{\bf T}^{2}=1} d^{2}{\bf T}~\mathcal{O}\left({\bf T}\right)P\left({\bf T}, {\bf T}_{0}, L\right).
\end{eqnarray}

One observable of interest is the mean square end-to-end distance $\left<\delta{\bf R}^2\right>$, which we compute in the same way as Sait$\hat{\rm o}$
\begin{equation}
 \left<\delta{\bf R}^2\right>=\int_{0}^{L}ds^{\prime}\int_{0}^{L}ds\left<{\bf T}(s)\cdot{\bf T}(s^{\prime})\right>.
 \end{equation}
 Particularly, when the polymer is lying in the open Euclidean plane $\left<\delta{\bf R}^2\right>$, the mean square end-to-end distance can be computed exactly and the result is the same as (\ref{planoabierto}). The argument to prove this is as follows. Due to the functional structure of the partition function, it is possible to prove that $\mathbb{Z}$ satisfies a diffusion-type equation $\frac{\partial \mathbb{Z}}{\partial s}=\frac{1}{2\ell_{p}}\nabla^{2}_{\bf T}\mathbb{Z}$, with the initial condition $\lim_{L\to 0}\mathbb{Z}\left({\bf T}_{L}, {\bf T}_{0}, L\right)=\delta\left({\bf T}_{L}-{\bf T}_{0}\right)$. Here the Laplacian operator $\nabla^2_{\bf T}$ is the two dimensional Laplacian constrained to ${\bf T}^2=1$, so it is convenient to define the parametrization ${\bf T}=\left(\cos\theta,\sin\theta\right)$. Thus, we have that
\begin{eqnarray}
\frac{\partial \mathbb{Z}}{\partial s}=\frac{1}{2\ell_{p}}\frac{\partial^2 \mathbb{Z}}{\partial\theta^2}.
\label{diff-eq}
\end{eqnarray}
The latter equation can be solved using separation of variables. The solution is 
\begin{eqnarray}
\mathbb{Z}\left(\theta, \theta_{0}, s\right)=\sum_{m=-\infty}^{\infty}e^{-\frac{m^2 s}{2\ell_{p}}}e^{im\left(\theta-\theta_{0}\right)}.
\label{Green-Plano}
\end{eqnarray}
Also, the normalization constant is $\mathcal{N}=2\pi$. Therefore, the mean square end-to-end distance reads
\begin{eqnarray}
\left<\delta{\bf R}^2\right>=\frac{1}{2\pi}\int_{0}^{L}ds^{\prime}\int_{0}^{L}ds \int_{0}^{2\pi} d\theta~\cos\theta ~\mathbb{Z}\left(\theta, 0, \left|s-s^{\prime}\right|\right).\nonumber\\
\end{eqnarray}
After an straightforward calculation using (\ref{Green-Plano}) one can conclude the desire result (\ref{planoabierto}). 

\section{Proof of Claims}\label{coefficients}

In this section we provide the proof of the assertion in claims \ref{result1} and \ref{result2}.

\setcounter{prop}{0}
\begin{prop} \label{result1}
Let $L/\ell_{p}$  any positive non-zero real number, then the mean square end-to-end distance \eqref{sol} obeys
\begin{eqnarray}
 \lim_{\ell_{p}/a\to 0}\frac{\left<\delta{\bf R}^2\right>_{\mathcal{D}}}{\ell^2_{p}}=\frac{4L}{\ell_{p}}-8\left(1-\exp\left(-\frac{L}{2\ell_{p}}\right)\right). \nonumber
\end{eqnarray}
\end{prop}
\begin{proof}  Expanding the function \eqref{Gfunction} in Taylor series in the variable $w$ at $w=0$ we find
\begin{equation}
G(v, w)=1-\frac{1}{2}\left(v-e^{-v}\sinh\left(v\right)\right)w+\mathcal{O}(w^2).
\label{EqG}
\end{equation}
Substituting \eqref{EqG} in Eq.~\eqref{sol}, we get
\begin{eqnarray}
\left<\delta{\bf R}^2\right>_{\mathcal{D}}&=&a^2\left[\frac{1}{3}-\sum_{n\in 2\mathbb{N}+1 }\frac{32}{\pi^4 n^4}\right]\nonumber\\&+&4\ell^2_{p}\left(\frac{L}{4\ell_{p}}-e^{-\frac{L}{4\ell_{p}}}\sinh\left(\frac{L}{4\ell_{p}}\right)\right)\sum_{n\in 2\mathbb{N}+1 }\frac{32}{\pi^2 n^2}\nonumber\\
&+&\mathcal{O}\left(\frac{\ell_{p}^3}{a^3}\right).
\label{inm}
\end{eqnarray}
Note that the series in \eqref{inm} can be expressed through the Riemann zeta and Dirichlet eta functions using
\begin{equation}
\sum_{n\in 2\mathbb{N}+1 }\frac{32}{\pi^k n^k}=\frac{16}{\pi^k}\left(\zeta\left(k\right)+\eta\left(k\right)\right).
\label{identidad}
\end{equation}
Specifically, for $k=2$ and $k=4$, we have  $\frac{16}{\pi^2}\left(\zeta\left(2\right)+\eta\left(2\right)\right)=4$, and $\frac{16}{\pi^4}\left(\zeta\left(4\right)+\eta\left(4\right)\right)=\frac{1}{3}$, respectively \cite{Gradshteyn}. Finally, we get the result claimed substituting these values in Eq.~\eqref{inm} and taking the corresponding limit.
\end{proof}

\begin{prop}\label{result2} Let $L/\ell_{p}$ and $\ell_{p}/a$ any positive non-zero real numbers and $c=2/3-64/\pi^4$, then the mean square end-to-end distance \eqref{sol} obeys 
\begin{eqnarray}
 0\leq \frac{\left<\delta{\bf R}^2\right>_{\mathcal{D}}}{a^2}\leq \frac{2}{3},\nonumber\end{eqnarray}
and
\begin{eqnarray}
 0\leq \frac{\left<\delta{\bf R}^2\right>_{\mathcal{D}}}{a^2}-\left(\frac{1}{3}-\frac{1}{3}G\left(\frac{L}{4\ell_{p}}, 8\pi^2\frac{\ell^2_{p}}{a^2}\right)\right)\leq c.\nonumber
\end{eqnarray}

\end{prop}

\begin{proof} For the first inequality, let us call \begin{eqnarray}
G_{n}\equiv G\left(\frac{L}{4\ell_{p}}, 8\pi^2\left(\frac{\ell_{p}}{a}\right)^2n^2\right),\label{gn}
\end{eqnarray}
where $G_{n}$ dependences on $L/\ell_{p}$ and $\ell_{p}/a$ are not written for the sake of simplicity. Taking the triangle inequality, and using that $\left|G_{n}\right|\leq 1$, we have that
\begin{equation}
\left|\sum_{n\in 2\mathbb{N}+1 }\frac{32}{\pi^4 n^4}G_{n}\right| \leq \sum_{n\in 2\mathbb{N}+1 }\frac{32}{\pi^4 n^4}=\frac{1}{3},
\label{ineq}
\end{equation}
where we have used the identity \eqref{identidad}, that is $\frac{16}{\pi^4}\left(\zeta\left(4\right)+\eta\left(4\right)\right)=\frac{1}{3}$. Finally, it is  sufficient to use the triangle inequality in Eq.~\eqref{sol} and the relation in Eq.~\eqref{ineq}.

For the second inequality, note that Eq.~\eqref{sol} can be written as follows
\begin{eqnarray}
\frac{\left<\delta{\bf R}^2\right>_{\mathcal{D}}}{a^2}-\left(\frac{1}{3}-\frac{1}{3}G_{1}\right)&=&\left(\frac{1}{3}-\frac{32}{\pi^4}\right)G_{1}\nonumber\\
&-&\sum_{n\in2\mathbb{N}+3}\frac{32}{\pi^4 n^4}G_{n}.
\end{eqnarray}
Taking the absolute value in both sides in the latter equation, and using the triangle inequality and the property $\left|G_{n}\right|\leq 1$, we have that
\begin{eqnarray}
\left|\frac{\left<\delta{\bf R}^2\right>_{\mathcal{D}}}{a^2}-\left(\frac{1}{3}-\frac{1}{3}G_{1}\right)\right|&\leq &\left(\frac{1}{3}-\frac{32}{\pi^4}\right)\nonumber\\
&+&\frac{16}{\pi^4}(\zeta(4)+\eta(4)-2).\nonumber
\end{eqnarray}
Using that $\zeta(4)=\pi^4/90$ and $\eta(4)=7\pi^4/720$, we conclude the proof.

\end{proof}


\end{document}